\documentclass[twocolumn]{aastex631}
\pdfoutput = 1

\usepackage{gensymb}
\usepackage{amsmath}

\begin{document}

\title{Expanding Mars Climate Modeling: Interpretable Machine Learning for Modeling MSL Relative Humidity}

\author{Nour Abdelmoneim}
\affiliation{Center for Space Science \\
New York University Abu Dhabi \\
PO Box 129188, Saadiyat Island, Abu Dhabi, UAE}

\author{Dattaraj B. Dhuri}
\affiliation{Center for Space Science \\
New York University Abu Dhabi \\
PO Box 129188, Saadiyat Island, Abu Dhabi, UAE}

\author{Dimitra Atri}
\affiliation{Center for Space Science \\
New York University Abu Dhabi \\
PO Box 129188, Saadiyat Island, Abu Dhabi, UAE}

\author{Germán Martínez}
\affiliation{Lunar and Planetary Institute \\
Universities Space Research Association \\
Houston, Texas 77058, USA}

\correspondingauthor{Nour Abdelmoneim}
\email{nour.abdelmoneim@nyu.edu}



\begin{abstract}

For the past several decades, numerous attempts have been made to model the climate of Mars with extensive studies focusing on the planet's dynamics and the understanding of its climate. While physical modeling and data assimilation approaches have made significant progress, uncertainties persist in comprehensively capturing and modeling the complexities of Martian climate. In this work, we propose a novel approach to Martian climate modeling by leveraging machine learning techniques that have shown remarkable success in Earth climate modeling. Our study presents a deep neural network designed to accurately model relative humidity in Gale Crater, as measured by NASA's Mars Science Laboratory ``Curiosity'' rover. By utilizing simulated meteorological variables produced by the Mars Planetary Climate Model, a robust Global Circulation Model, our model accurately predicts relative humidity with a mean error of 3\% and an $R^2$ score of 0.92. Furthermore, we present an approach to predict quantile ranges of relative humidity, catering to applications that require a range of values. To address the challenge of interpretability associated with machine learning models, we utilize an interpretable model architecture and conduct an in-depth analysis of its internal mechanisms and decision making processes. We find that our neural network can effectively model relative humidity at Gale crater using a few meteorological variables, with the monthly mean surface H$_2$O layer, planetary boundary layer height, convective wind speed, and solar zenith angle being the primary contributors to the model predictions. The emphasis on interpretability significantly improves the reliability of this approach, making it suitable for Martian scientific research. In addition to providing a fast and efficient method to modeling climate variables on Mars, this modeling approach can also be used to expand on current datasets by filling spatial and temporal gaps in observations.
\end{abstract}



\section{Introduction} \label{sec:intro}

Findings from decades of Mars exploration suggest that around 3.5 billion years ago, Mars had a wet environment that could have been suitable for sustaining life. Research on the Martian surface has uncovered evidence of ancient oceans that existed on the planet \citep{baker1991ancient, head1999possible}. NASA's Mars Science Laboratory (MSL) ``Curiosity'' rover discovered evidence of an ancient lake with an aqueous environment that had the necessary characteristics to support a martian biosphere, such as low salinity and neutral pH \citep{grotzinger2014habitable}. A recent study reports observations of polygonal structures from MSL that show evidence of seasonal wet-dry cycling in early Mars, suggesting an Earth-like wet climate \citep{rapin2023sustained}. Evidence of geological features that have been altered by liquid water over 2 billion years ago has also been discovered in Jezero crater by NASA's Mars 2020 ``Perseverance'' rover \citep{scheller2022aqueous, farley2022aqueously, hamran2022ground}. Since water is a requirement for life as we know it, the search for water on present-day Mars continues to be an active area of research. Water ice on the surface of the planet has been found in the north and south poles \citep{titus2003exposed, bibring2004perennial} as well as in the regolith at other locations \citep{mellon1997persistence, audouard2014water}. While harsh conditions such as the planet's thin atmosphere and low temperatures \citep{atri2023diurnal} make it difficult for liquid water to exist on the surface of the planet, high relative humidity (RH) is an indicator of the presence of subsurface water reservoirs \citep{martinez2013water, rivera2018constraining}. Studies suggest that transient liquid water could exist at Gale crater under certain temperature and RH conditions in addition to the presence of perchlorate salts that lower the freezing temperature of water \citep{martin2015transient}. Furthermore, other experimental studies have suggested that liquid brines can form in the shallow subsurface of Mars in the presence of perchlorates and water ice \citep{fischer2014experimental, gough2023laboratory}. Additionally, experimental investigations have shown that liquid water can exist on the surface of Mars under certain conditions when RH levels are near saturation \citep{vakkada2021experimental, nikolakakos2018laboratory}. Since RH is closely linked to the presence or absence of water on the planet's surface, understanding and modeling RH is critical and has potential implications for habitability. 

Several sophisticated physical models have been developed to simulate Martian climate. One of the first climate models developed was the NASA Ames' General Circulation Model (GCM) that began as a simpler model assuming a dust-free, pure CO$_2$ atmosphere \citep{pollack1981martian}. Since then, Mars GCMs have developed into complex models that can reasonably simulate Martian conditions. One notable model is the Planetary Weather Research and Forecasting model (planetWRF) which is a modified version of the terrestrial Weather Research and Forecasting (WRF) model \citep{richardson2007planetwrf}. PlanetWRF models the atmosphere of several planetary objects including Mars and Titan with varying resolutions. In this paper, we will be using the Mars Planetary Climate Model (PCM), formerly known as the Laboratoire de Météorologie Dynamique Mars Global Climate Model (LMD Mars GCM) \citep{forget1999improved}. The Mars PCM is a sophisticated GCM that models the atmosphere and climate of Mars in 3D, simulating the water cycle, dust transport, and atmospheric compositions. More details on the Mars PCM and its derived Mars Climate Database (MCD) can be found in Section \ref{sec:mcd_data}.

While numerical models have been extensively developed to model Martian climate, the potential application of artificial intelligence (AI) methods to model the climate of Mars has yet to be extensively explored in scientific research. To our knowledge, two machine learning studies have been carried out on Martian climate, both on maximum/minimum daily temperature time series forecasting tasks. \citet{priyadarshini2021mars} use NASA's Curiosity rover data to predict maximum daily temperature from terrestrial date as input. They compare several machine learning algorithms concluding that a stacked Long Short-Term Memory (LSTM) network, a type of neural network that can learn temporal dependencies, performs best in predicting maximum daily temperatures. Another attempt uses Curiosity data to predict maximum and minimum daily temperatures \citep{al2022study}. They run 4 experiments: two experiments use terrestrial date as input to the machine learning model, one predicts maximum daily temperature and the second minimum daily temperature, while the two other experiments use sol number as input to also predict maximum/minimum daily temperatures. 

While machine learning studies on Martian climate are limited, machine learning has been utilized extensively for modeling Earth climate. Using machine learning in the field of climate science has improved our understanding of Earth's climate and its interactions with other systems. One effective application of machine learning in weather prediction is a model presented by \citet{pathak2022fourcastnet} called FourCastNet. This model is trained on a comprehensive climate dataset (ERA5) \citep{hersbach2020era5} that consists of over 50 years of hourly estimates of different meteorological features. ERA5 is constructed by assimilating observations and numerical models at a high spatial resolution of $0.25\degree \times 0.25\degree$. Using this dataset, FourCastNet is successfully trained to forecast challenging climate features such as wind speeds and precipitation at 6-hour intervals making it capable of identifying extreme weather events including hurricanes. Additionally, it outperforms the state of the art numerical weather prediction models in terms of computational efficiency, being 45,000 times faster at making inferences. Machine learning has been evaluated on RH prediction tasks in several studies focusing on different locations on Earth. In a study on the Terengganu state in Malaysia, several algorithms including linear regression, tree-based algorithms, and neural network architectures are trained to forecast daily and monthly RH given RH measurements for several days/months prior to the forecast day/month \citep{hanoon2021developing}. Their findings suggest that neural networks are more effective than other algorithms in predicting RH. Similar studies have been carried out on RH in different locations on Earth, also coming to the conclusion that neural networks were effective at predicting RH \citep{ozbek2022daily, shad2022forecasting}. In addition to models that are purely based on AI, hybrid approaches combining numerical models and AI, referred to as Neural Earth System Modeling (NESYM), have been used to model Earth's climate \citep{irrgang2021towards}. Weakly coupled NESYM is a branch of NESYM that allows information flow between physical models and AI. Not only does this approach decrease the amount of data required for training a machine learning model, but it also improves the physical consistency of the model outputs. 

With the availability of advanced physical models of Martian climate along with an increasing amount of data from various missions to Mars, we explore a hybrid approach to climate modeling by combining AI with existing physical models of the Martian climate. In this paper, we present an interpretable machine learning model that effectively models the RH near the Martian surface using meteorological variables obtained from existing physical modeling. We demonstrate that machine learning can be used to expand on existing Mars climate modeling techniques because of its ability to capture complex nonlinear relationships between variables. We also show that machine learning models can be designed to be interpretable while being fast and accurate, making it a useful tool for analyzing Martian climate data. One practical application of this modeling approach is its ability to produce accurate synthetic data at the rover's location, effectively filling spatial and temporal gaps in observations and expanding data coverage. Furthermore, our results show that RH levels on Mars can be modeled using a small set of numerically modeled meteorological variables that are not conventionally used when physically modeling humidity; this finding opens the possibility of alternative modeling techniques. 

The paper is structured as follows: Section \ref{sec:msl_data} and \ref{sec:mcd_data} describe the MSL and Mars PCM training data used to develop our model. Section \ref{sec:data_prep} outlines our feature selection and data preparation process. Section \ref{sec:methods} describes the model architecture with an explanation of its interpretable nature (Section \ref{sec:mlmodel}), the objective functions used to train the model (Section \ref{sec:loss_functions}), the hyperparameter tuning process (Section \ref{sec:hyperparam}), and the baseline models used for evaluation (Section \ref{sec:baselinemethods}). Our results are discussed in Section \ref{sec:results}; Section \ref{sec:performance_results} discusses the performance of our trained models and Section \ref{sec:interpretation_results} discusses the interpretation of the model's outcome. Lastly, a discussion of our results, the scope of our analysis, and potential future improvements can be found in Section \ref{sec:discussion}.

\section{Data}
\label{sec:data}
\begin{figure*}
\centering
	\includegraphics[width=1.0\linewidth]{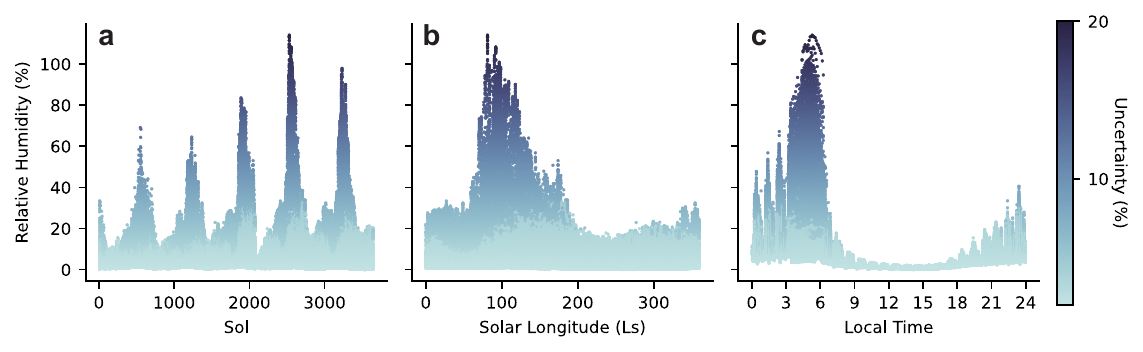}
    \caption{\textbf{Relative humidity observations by REMS.} Panel \textbf{a} shows 5 years of REMS RH. Seasonal variation is shown in panel \textbf{b}. Diurnal variation is shown in panel \textbf{c}.}
    \label{fig:humidity-data}
\end{figure*}

\subsection{MSL Relative Humidity}
\label{sec:msl_data} 

The MSL Curiosity rover landed at Gale crater (4.5\degree\ south latitude, 137.4\degree\ east longitude) on August 5, 2012 and has been operational for over 5 Mars years. The Rover Environmental Monitoring Station (REMS) on board MSL is a weather station that provides meteorological measurements around the rover including surface and air temperatures, pressure, RH, winds, and ultraviolet radiation \citep{sebastian2010rover, gomez2012rems}. The Relative Humidity Sensor (RHS), mounted on one of the REMS booms at 1.6 meters above the ground, measures RH within a range of 0-100\% \citep{harri2014mars}. It has an overall accuracy of around 10\%, however, its accuracy varies based on ambient temperature. Due to self-heating of the RHS, reliable RH measurements include only those taken within the first 4 seconds after the RHS has been turned off for at least five minutes \citep{martinez2017modern}. These measurements include those taken during the nominal and the high-resolution interval mode (HRIM), which consists of switching the sensor on and off at periodic intervals to minimize heating. We take the average of each 4-second interval of measurements resulting in 75,839 samples from Curiosity sols 9 to 3644 spanning over 5 Mars years (MY31 to MY36). 

REMS RH measurements show strong diurnal and seasonal patterns as displayed in Figure \ref{fig:humidity-data}. RH peaks around sunrise, when the air temperature is coldest. It subsequently drops dramatically during daytime as air temperature increases, reaching minimum values between 0\%-5\% in the afternoons when air temperature is at its warmest. RH at the MSL site also varies seasonally; higher RH levels (up to 100\%) are observed during the winter season in the southern hemisphere (L$_{s}$ = 90-150). Figure \ref{fig:humidity-data} also shows the calibration uncertainty of the data. Calibration uncertainty ranges from 0\% to 20\% RH and is significantly higher for higher RH values. 

To develop our machine learning model, the first 60\% of data is used for training which approximately covers 3 Mars years, the following 20\% (1 Martian year) for validation, and the final 20\% for testing. Machine learning models are data-driven and are therefore very sensitive to the distribution of training data that they receive. RH levels at Gale Crater are generally low with over 84\% of the samples being below 20\% RH. To balance the training set, we divided the training dataset into 5 bins: 0-20\% RH, 20-40\% RH, 40-60\% RH, 60-80\%, and 80-100\% RH. Then, we randomly sample 10,000 data points from each bin. Since the first bin (0-20\% RH) contained over 10,000 samples, we under-sampled the data within this range. However, the remaining bins contained less than 10,000 samples each; therefore, we randomly over-sampled the data points in each of these ranges. By adopting this approach, the model is trained on a balanced dataset and does not learn to optimize its performance to predict only low RH values. 

\subsection{Mars Climate Database}
\label{sec:mcd_data}

The Mars Climate Database is derived from the Mars PCM, a global climate model of Mars that is based on numerical simulations and validated with observations \citep{lewis1999climate, millour2019latest}. The Mars PCM computes a 3D atmospheric and climate model that includes dust \citep{madeleine2011revisiting} and water cycle \citep{navarro2014global} simulations. The MCD contains information about atmospheric variables such as temperature, pressure, wind, and density, amongst 85 other features. Throughout this paper, we use the term ``feature'' to refer to ``variable'' as this is the terminology commonly used to refer to machine learning model input variables. Before proceeding with the analysis of MCD features, we manually eliminate features that would not provide useful information to our neural network, such as features that remain constant and ones relating to the day-to-day RMS variability of certain quantities. The remaining MCD output features are listed in Figure \ref{fig:mi_results}. To accurately simulate local pressure and density, the MCD uses MOLA topography (32 pixels/ degree) for high resolution interpolation. We obtain all the MCD outputs at the specific times and locations of the Curiosity rover observations. This forms our input data for predicting the corresponding RH measurements. 

\subsection{Data Processing and Feature Selection}
\label{sec:data_prep}

Feature selection improves the performance and predictive capabilities of machine learning algorithms by eliminating redundant and/or irrelevant data while allowing the model to learn more meaningful representations of the data \citep{Cai2018}. We use mutual information estimation, a filtering feature selection method, to narrow down the features computed by the MCD. Mutual information, often known as ``information gain'', is a well-established measure of mutual dependence between two features in information theory \citep{thomas1991elements, gray2011entropy}. It is a measure of the reduction in entropy of a certain variable in the presence of another. Therefore, high mutual information indicates that two variables contain useful information about one another, whereas independent variables would have zero mutual information.

Mutual information for two continuous variables is defined as
\begin{equation}
    I\left(X;Y\right) = h\left(X\right) - h\left(X|Y\right),
    \label{eq:mi}
\end{equation}
where $h(X)$ is the differential entropy for a continuous random variable $X$ with a probability density function $f(x)$; $h(X)$ is defined as
\begin{equation}
    h\left(X\right) = -\int{f\left(x\right)\log{f\left(x\right)}dx},
    \label{eq:entropy1}
\end{equation}
and $h(X|Y)$ is the conditional entropy between two continuous random variables $X$ and $Y$ defined as
\begin{equation}
    h\left(X|Y\right) = -\iint{f\left(x,y\right)\log{f\left(x|y\right)}dxdy},
    \label{eq:entropy2}
\end{equation}
where $f(x,y)$ and $f(x|y)$ are the joint and conditional probability density functions.

We compute mutual information for each MCD feature with RH to identify features that are relevant to the RH modeling task. Computing mutual information for discrete variables is straightforward, however all of the variables in our study are continuous. Since the marginal probability densities of the variables are unknown, an approximation algorithm is necessary to estimate the mutual information between continuous variables. A common method used to estimate mutual information is discretizing the continuous variables into bins with \textbf{n} points each, however this approach has been shown to be highly sensitive to the value of \textbf{n} and can result in inaccurate estimations of mutual information \citep{ross2014mutual}. Instead, we use the k-Nearest Neighbor (k-NN) algorithm described in \citep{kraskov2004estimating, ross2014mutual}. Comparisons between the two methods presented by \citet{ross2014mutual} show that the k-NN method is more robust resulting in more accurate estimations of mutual information. It is recommended that the value of \textbf{k} is a small integer in the range of 1-10, we set \textbf{k} = 9 due to our large dataset. 

\begin{figure}
\centering
	\includegraphics[width = 1.0\linewidth]{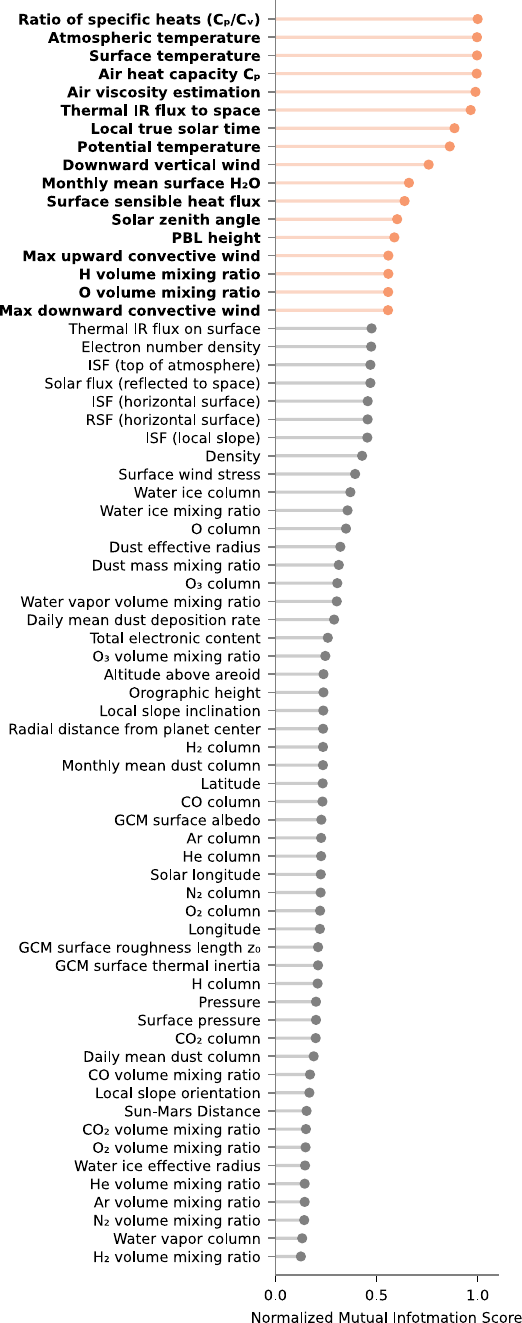}
    \caption{Normalized mutual information scores of MCD features with REMS RH. Features with normalized mutual information scores of $>=$ 0.5 are highlighted.}
    \label{fig:mi_results}
\end{figure}

After computing mutual information scores between each input feature and the target variable (RH), scores are normalized between 0 and 1 by dividing by the maximum mutual information. The normalized scores are shown in Figure \ref{fig:mi_results}. In order to reduce the number of inputs to the neural network, we apply a threshold on mutual information scores and only consider features with scores above the threshold for training our machine learning model. Since mutual information measures the amount of information a certain variable conveys about another, we filter out features with normalized mutual information scores below 0.5. This results in a more consistent training process for the neural network which makes its interpretation clearer and more reliable. Among the 69 remaining MCD output features, 17 had normalized mutual information scores greater than or equal to 0.5 (highlighted in orange/bold in Figure \ref{fig:mi_results}). These included variables such as wind speeds and surface and atmospheric temperatures, in addition to variables related to atmospheric constituents. This leaves us with a set of features that provide a good amount of information about RH and eliminates irrelevant features that would slow down the training and inference processes and complicate the interpretation of the model.

To further reduce the number of features used to train the model, we use a feature clustering approach based on Pearson correlation coefficients. Specifically, clusters are formed with features that have pairwise correlations that are greater than or equal to 0.98. Then, within each cluster, the feature with the highest mutual information score is selected to train the model. Features that do not belong to any cluster are also included. The algorithm used for clustering is feature agglomeration and the feature clusters are shown in Table \ref{tab:cluster_table}. The rationale behind this strategy is that, in theory, the variable selection network can extract the same information from features that are highly correlated. Thus, providing the model with highly correlated features is redundant and may complicate the model's interpretability while having a detrimental effect on its performance. Additionally, feature clusters are semantically related so this approach reduces the complexity of the feature space and improves our semantic understanding of the importance of different features to the neural network. The selected input features after applying the aforementioned feature selection steps are listed in Figure \ref{fig:vsn-grn}.

\begin{deluxetable}{lll}
\tablecolumns{3}
\tablecaption{Input feature clusters}
\tablehead{\textbf{Group} & \textbf{Feature} & \textbf{Average} \\
 ~ & ~ & \textbf{correlation}}
\startdata
    \textbf{Convective} & \textbf{Max upward} & 1.0  \\
    \textbf{wind} & \textbf{convective wind} & ~ \\
    ~ & \textbf{within the PBL}\\ 
    ~ & Max downward & ~ \\ 
    ~ & convective wind & ~ \\
    ~ & within the PBL & ~ \\ 
    \hline
    \textbf{Temperature} & \textbf{Atmospheric} & 0.99 \\ 
    ~ & \textbf{temperature} & ~ \\
    ~ & Surface temperature & ~ \\ 
    ~ & Potential temperature & ~ \\ 
    ~ & Thermal IR flux to space & ~ \\ 
    ~ & Air heat capacity & ~ \\ 
    ~ & Air viscosity estimation & ~ \\ 
\enddata
\tablecomments{Input feature clustered by Pearson correlation coefficients. All features in each group have pairwise correlations of $>=$ 0.98. Additional features not belonging to clusters are not listed. Selected features from each group are highlighted.}
\label{tab:cluster_table}
\end{deluxetable}

\section{Methods}
\label{sec:methods}

\subsection{Model Architecture and Interpretability}
\label{sec:mlmodel}

\begin{figure*}
    \centering
	\includegraphics[width=17cm]{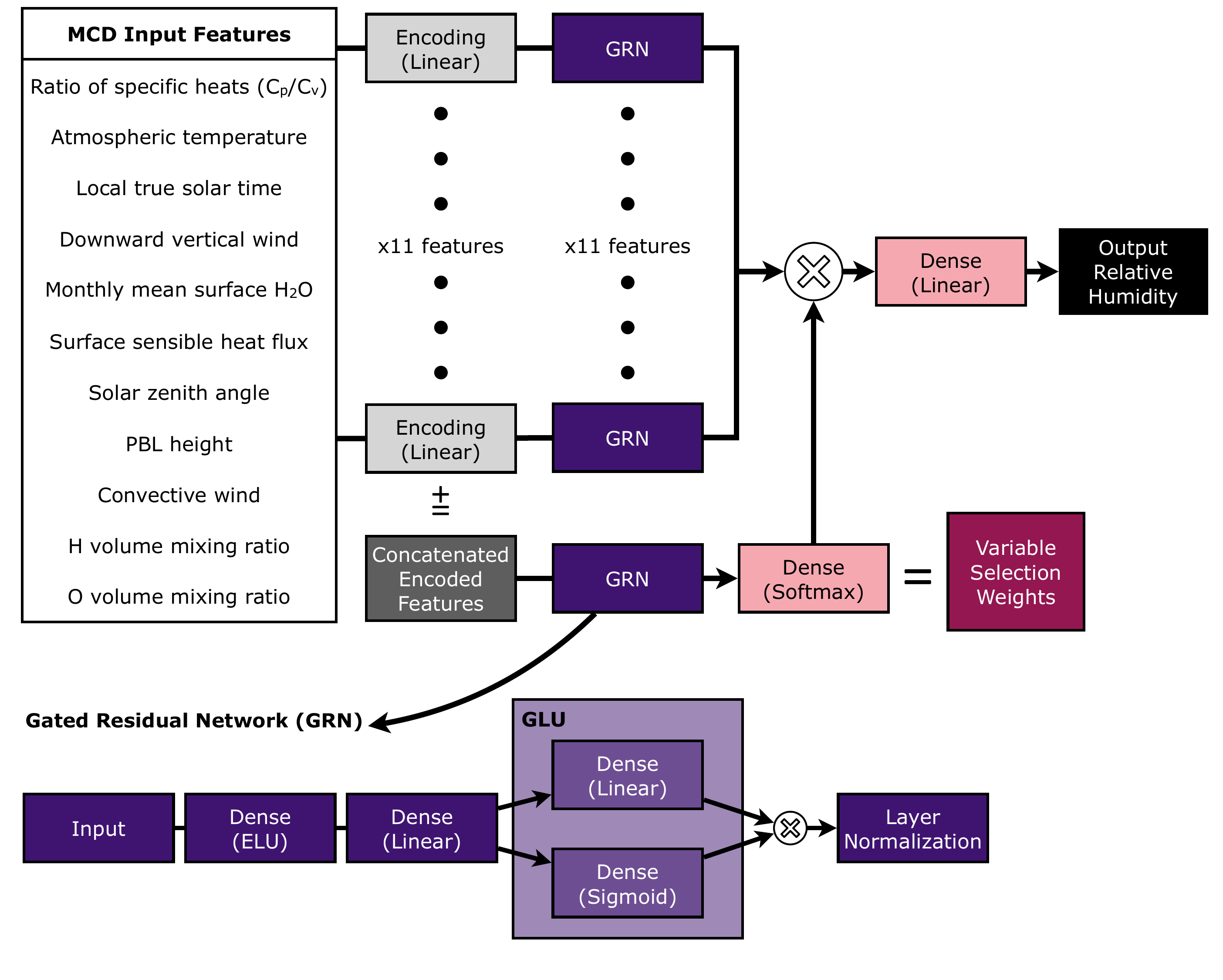}
    \caption{Neural network architecture. The VSN consists of the layers past the encoding layers and prior to the output dense layer. The model receives 11 MCD features as inputs. Each feature is then linearly encoded and fed individually into its own GRN. The 11 encoded MCD features are also concatenated and fed into one additional GRN that outputs the variable selection weight for each feature. The GRN architecture contains a GLU which determines whether the network will include an additional non-linear (sigmoid) layer. The output of each individual feature GRN is then weighted with its variable selection weight and combined to form the RH prediction. Dense layers, or fully connected layers, form the building blocks of this architecture. For each dense layer, the activation function used is indicated in the corresponding parentheses.}
    \label{fig:vsn-grn}
\end{figure*}

Since we are training our model with a large number of input features that have varying relationships with RH, we use a variable selection network (VSN) proposed by \citet{lim2021temporal} to discern the salient features. The VSN computes variable selection weights that are applied to each feature to produce the output RH prediction. The VSN includes a gated residual network (GRN) which controls the complexity of the model by only using non-linear processing when necessary. This allows the model to have adaptable complexity that can adjust according to the input data. The GRN utilizes gated linear units (GLUs) to suppress nonessential layers from the model architecture. Figure \ref{fig:vsn-grn} shows the architecture of the VSN, GRN, and GLU with more details on the information flow between the different layers of the model.

In comparison with black-box machine learning models which require post-hoc methods to explain, this model provides inherent interpretability via the VSN. This network receives all the input features and passes them through a dense layer with a softmax activation function:
\begin{equation}
    {Softmax(x_i) = \frac{e^{x_i}}{\sum_{j=1}^{K}{e^{x_j}}}}.
    \label{eq:softmax}
\end{equation}
The dense layer has trainable parameters that are adjusted throughout the training process to produce more accurate RH predictions. By applying the softmax activation function, each output of the layer is transformed to lie in the range [0-1] (with all weights summing to 1) representing the weight each feature is assigned. Consequently, features that are assigned higher weights by the VSN contribute more to the predicted RH. By examining the results of the VSN, we can see the importance of every input feature to the corresponding RH output for each sample. Using this, we can examine the magnitude of each feature's contribution to the model output. 

\subsection{Objective Functions}
\label{sec:loss_functions}

In order to train a neural network, an optimization process is required that relies on an objective function (or ``loss'' function) to evaluate the network's performance. The choice of loss function determines the type of errors the model is trained to minimize. We present results using different loss functions in the training process and compare the performance of each model. For predicting continuous variables (in our case, RH), the most basic loss function is the mean absolute error ($MAE$) which is defined as
\begin{equation}
    MAE\left(y, \hat{y}\right) = \frac{ \sum_{i=1}^{N} |y_i - \hat{y}_i|}{N}, 
    \label{eq:mae}
\end{equation}
where y is the ground truth, $\hat{y}$ is the model prediction, and N is the number of samples being evaluated. The $MAE$ loss weighs all errors and samples equally, with the goal of minimizing the average error across the model. While models trained with $MAE$ loss can effectively minimize the majority of errors, they may also end up with some very large errors for a few number of samples. The mean squared error ($MSE$) loss helps mitigate this issue by giving more weight to larger errors. Models trained with $MSE$ loss prioritize minimizing larger errors more than smaller errors, since the error of each sample is squared. $MSE$ is defined as
\begin{equation}
    MSE\left(y, \hat{y}\right) = \frac{\sum_{i=1}^{N}\left(y_i - \hat{y}_i\right)^2}{N}.
    \label{eq:mse}
\end{equation}

\begin{figure}
\centering
	\includegraphics[width = 1.0\linewidth]{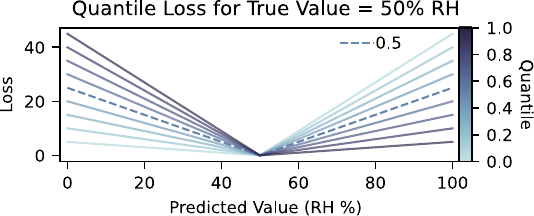}
    \caption{Behavior of quantile loss. At a true RH value of 50\%, quantile loss is larger for model predictions that are $<$50\% for higher quantiles, effectively penalizing under-predictions more than over-predictions. The opposite is true for lower quantiles that penalize over-predictions more. Quantile 0.5 is highlighted with a dashed line, where the loss is the same for over and under-predictions, making it equivalent to $MAE$ in terms of optimization.}
    \label{fig:qloss_plot}
\end{figure}

For some applications, predicting a range of values can be more useful than a single value. Deep quantile regression is a form of deep learning that outputs prediction intervals instead of point estimates. A prediction interval allows us to report a range of values along with a certain level of confidence that the true value lies within this range. Unlike previous loss functions that aim to minimize the error between the ground truth and predicted values, quantile loss is designed to minimize the error at specific quantiles of the target distribution. This is particularly useful when modeling unbalanced data with outliers or skewed distributions, which is the case with our RH data. In fact, 65\% of the data falls within the 0\%-10\% range and only 1.5\% of RH measurements fall above 70\% RH. The quantile loss function is defined as 
\begin{equation} 
    {qloss}\left(y, \hat{y}, q\right) = max[q\left(y_i - \hat{y}_i\right), \left(q-1\right)\left(y_i - \hat{y}_i\right)], 
    \label{eq:qloss}
\end{equation}
where q is the quantile being predicted. It is important to note that when q=0.5, the quantile loss only differs from $MAE$ by a constant, making them equivalent for optimization purposes. Figure \ref{fig:qloss_plot} shows the behavior of the quantile loss function for different quantiles.

\subsection{Hyperparameter Tuning}
\label{sec:hyperparam}

Machine learning models are trained with a certain set of parameters called hyperparameters that require tuning. These include the batch size, learning rate, dropout rate, and encoding size. The batch size is the amount of data the model processes before updating its parameters during training. Neural networks are trained using stochastic gradient descent (SGD), thus the learning rate determines the step size at each iteration during the SGD optimization process. The dropout rate is the percentage of neurons that are randomly dropped at every training iteration. Dropping neurons can help prevent overfitting by ensuring that the nodes of the model are not co-dependant on one another. The final hyperparameter we tune is encoding size, which is specific to our model architecture. This is the dimension of the encoding layer in the neural network that each input feature is passed through. To tune these hyperparameters, we define a range of values for each hyperparameter and use grid search to train models with all possible combinations of hyperparameter values. Table \ref{tab:hyperparam_table} displays the hyperparameter values that were tested. We then use 5-fold cross validation for each combination of hyperparameters, that is, we divide the dataset into 5 portions (``folds'') and train the model with the same combination of hyperparameters 5 times, each iteration using 4 folds for testing and 1 fold for validation. Following this, we take the average validation MSE loss for each combination of hyperparameters and select the hyperparameters that result in the lowest MSE loss. 

\begin{deluxetable}{lllll}
    \tablewidth{1.0\columnwidth}
    \tablecolumns{5}
    \tablecaption{Hyperparameter values. 
    \label{tab:hyperparam_table}}
    \tablehead{\textbf{Hyperparameter} & \textbf{Values}}
    \startdata
        Batch Size & 16 & 32 & \textbf{64}\\
        Learning Rate & \textbf{0.001} & 0.0001 & 0.00001\\
        Dropout Rate & \textbf{0.0} & 0.1 & 0.2\\
        Encoding Size & 32 & 64 & \textbf{128}\\
    \enddata
    \tablecomments{The range of hyperparameter values used in grid search. Best performing hyperparameters are highlighted.}
\end{deluxetable}

\subsection{Baseline Models}
\label{sec:baselinemethods}

We use two baseline models to compare the performance of our proposed neural network architecture: a multiple linear regression model and a basic neural network. Both baseline models were trained with the same set of features obtained from the feature selection steps above. This was done to ensure a more accurate comparison between our proposed neural network architecture and the baseline models without the influence of the feature selection preprocessing step. We also use the same train/validation/test split that we employed to train our proposed neural network architecture. Our first baseline model is a multiple linear regression model. Multiple linear regression is a statistical modeling technique that is used to estimate the relationship between one dependent variable (RH) and multiple independent variables (MCD features) by fitting a linear model that minimizes the residual sum of squares between the output predictions of the linear approximation and the ground truth data. Our second baseline model is a neural network with a very basic architecture compared to the one we present in this work. The baseline neural network consists of 5 layers in total: an input layer, an output layer, and 3 dense hidden layers in between. A dropout layer was inserted after every hidden layer to avoid overfitting. We apply the same hyperparameter tuning process with grid search to find the optimal hyperparameters for this baseline neural network. The ``encoding size'' hyperparameter that was only applicable to our proposed deep neural network architecture was replaced with another hyperparameter that is relevant to the baseline architecture: multilayer perceptron (MLP) dimension. The MLP dimension determines the output dimension of each of the hidden dense layers in the neural network. The remaining hyperparameters (batch size, learning rate, and dropout rate) are directly applicable to the baseline neural network. The results of the baseline models are presented in section~\ref{sec:performance_results} along with the performance of the deep neural network architecture.

\section{Results}
\label{sec:results}


\subsection{Performance}
\label{sec:performance_results}

Table \ref{tab:result_table} shows the baseline models' scores on different performance metrics compared to those of our deep neural network. Our neural network outperforms baseline models on the MSE, MAE, R (Pearson correlation), and $R^2$ (coefficient of determination) metrics with scores of 24.23, 3.03, 0.96, and 0.92 respectively. The $R^2$ score indicates high correlation between the model predictions and true humidity values measured by the Curiosity rover.

\begin{deluxetable}{lcccc}[h!]
    \tablecolumns{5}
    \tablecaption{Model Performance Metrics}
    \tablehead{Model & $MSE$ & $MAE$ & $R$ & $R^2$}
    \startdata
          Multiple linear regression & 102.44 & 5.28 & 0.85 & 0.65\\
          Simple neural network & 33.92 & 3.61 & 0.95 & 0.88\\
          Deep neural network (VSN) & \textbf{24.23} & \textbf{3.03} & \textbf{0.96} & \textbf{0.92}\\
    \enddata
    \tablecomments{Model performance results compared to baseline models. Best performing metrics are highlighted in bold. $MAE$ and $MSE$ are expressed in \%RH and \%RH$^2$ respectively. Lower $MSE$ and $MAE$ indicate less error in predictions, while higher $R$ and $R^2$ values indicate better predictions.}
    \label{tab:result_table}
\end{deluxetable}

\begin{figure}
\centering
	\includegraphics[width = 1.0\linewidth]{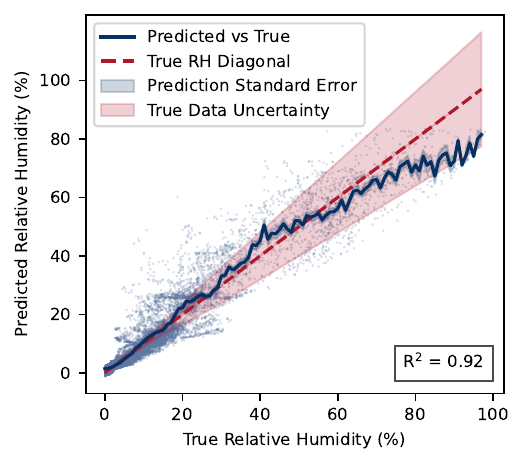}
    \caption{Model predictions compared with ground truth data for the 15,158 samples in the test set. The blue solid line shows the model predictions vs. the true RH binned by 1\% RH; one standard error is indicated with the blue shaded region. A perfect outcome is shown (red dashed diagonal), with the calibration uncertainty range of the REMS RH data highlighted in the red shaded region.}
    \label{fig:error_uncertainty_plot}
\end{figure}

Figure \ref{fig:error_uncertainty_plot} shows the performance of the deep neural network across the range of RH. Additionally, we show the calibration uncertainty range of the REMS measurements, which ranges from  0\% to 20\% uncertainty and increases linearly with RH. The deep neural network achieves good performance in modeling RH up to 80\% since the majority of the model predictions lie within the range of the REMS instrument uncertainty. Its performance deteriorates as RH levels rise above 80\% as the model consistently under-predicts. Despite the model's under-predictions past 80\% RH, most of the RH predictions remain within the instrument uncertainty. The decrease in performance can be attributed to the limited number of samples at higher RH levels since the performance of machine learning models heavily depends on the availability of data. 

\begin{figure}
\centering
	\includegraphics[width = 1.0\linewidth]{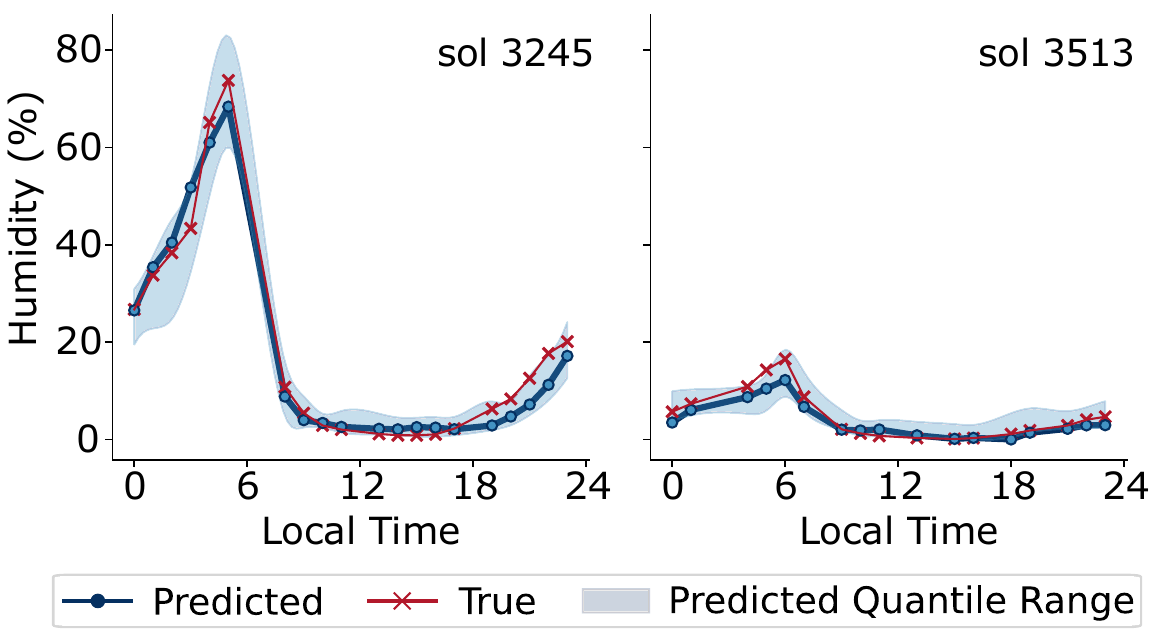}
    \caption{Predicted and observed (true) RH for sols with high and low RH values. Solid blue lines show the predicted RH and blue shaded regions show the predicted quantile range (P10 and P90). The true RH values are shown in red.}
    \label{fig:result_sols}
\end{figure}

In addition to the model trained with MSE loss, we train two models using quantile loss to predict the 0.1 quantile (P10) and 0.9 quantile (P90) values. The quantile losses for P10 and P90 are 0.67 and 0.65 respectively. Ideally, 80\% of the RH measurements should lie within the predicted P10 and P90 quantile range. Our results from the 15,158 samples in the test dataset show that 78\% of the measurements lie within the predicted quantile range. We show two examples of our model predictions and quantile outcomes in Figure \ref{fig:result_sols}: one example of a sol with high RH (sol 3245) and one with low RH (sol 3513). 

\subsection{Interpretation}
\label{sec:interpretation_results}

\begin{figure*}
\centering
	\includegraphics[width=1.0\linewidth]{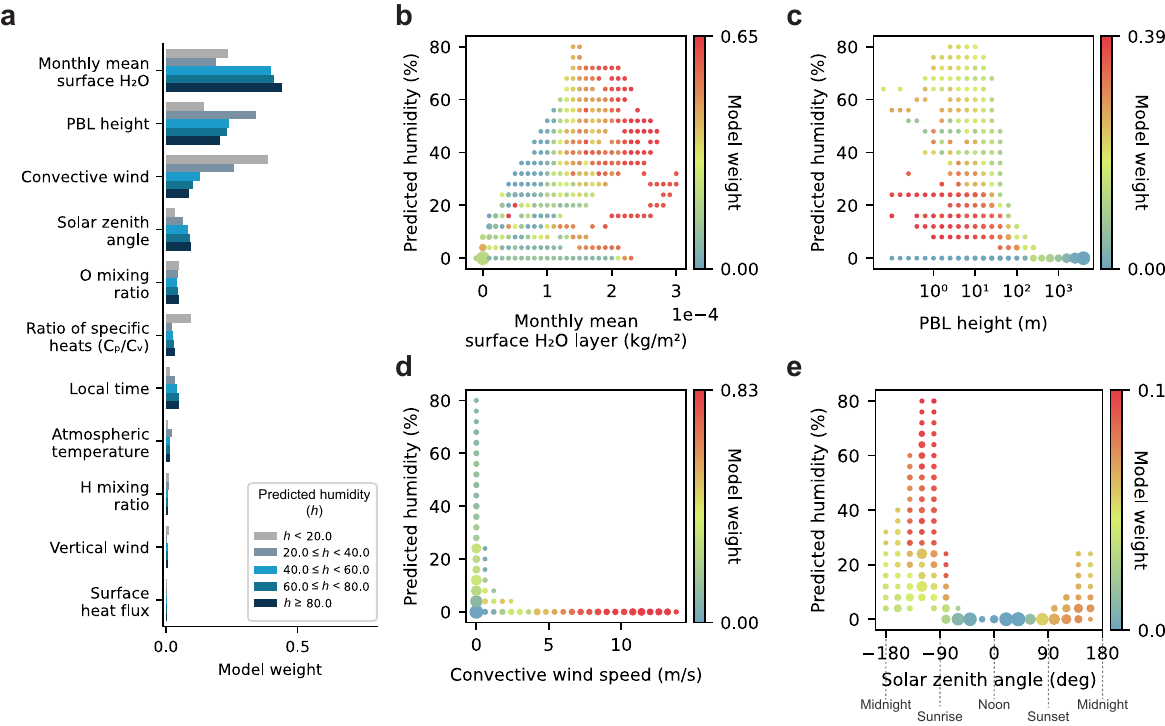}
    \caption{Variable Selection Network results. Panel \textbf{a} shows the average weights assigned by the variable selection network (model weights) for each input feature grouped by predicted humidity ranges. Panels \textbf{b}, \textbf{c}, \textbf{d}, and \textbf{e}, display the model weights of the top 4 input features. They also show how model weights vary with each respective feature values and predicted humidity. Circle sizes represent the number of samples present.}
    \label{fig:vsn-results}
\end{figure*}

The neural network architecture used in our model allows us to examine which input features are being considered by the model for each RH prediction and the extent to which each input feature contributes to the output. For each sample in our test set, the model receives 11 inputs (MCD features); our interpretation results indicate the weight that each feature is attributed by the model for each sample. In effect, we can examine the input features at play for each RH inference that the machine learning model makes. 

Figure \ref{fig:vsn-results} summarizes key results from our model interpretation. Figure \ref{fig:vsn-results}\textbf{a} depicts the model weights assigned to each of the input features across the 15,158 samples in the test set grouped by predicted RH range. As the model receives 11 input features, any weight exceeding 1/11 or approximately 0.091 denotes that the feature is significantly considered by the model. Figures \ref{fig:vsn-results}\textbf{b}-\textbf{e} illustrate how the feature weights for the 4 highest weighted inputs vary for each test sample. Given the capability of machine learning models to learn complex relationships between variables, the weights our model assigns to each input feature vary considerably across different samples, albeit still exhibiting patterns that have physical meaning. 

Our analysis demonstrates that the monthly mean surface H$_2$O layer input feature carries significant weight in our model across all five RH groups. The monthly mean surface H$_2$O layer is a monthly average of the amount of non-perennial frost in a certain area. Its impact on the output RH prediction is more pronounced within higher RH groups (40-100\% RH) compared to lower RH groups (0-40\% RH). This trend is also demonstrated in Figure \ref{fig:vsn-results}\textbf{b}, which illustrates how the model weights for surface H$_2$O vary in response to changes in both RH levels and the size of the surface H$_2$O layer. Furthermore, it appears that the model places greater weight on this feature as its values increase, suggesting that substantial information regarding RH can be drawn from elevated values of the surface H$_2$O layer. 

Figure \ref{fig:vsn-results}\textbf{c} illustrates the model weights for the PBL height input feature. The PBL height varies diurnally with lower PBL heights occurring at night time and higher PBL heights at day time. As demonstrated in Figures \ref{fig:vsn-results}\textbf{a} and \ref{fig:vsn-results}\textbf{c}, the model heavily relies on PBL height when predicting RH ranges from 20 to 40\% which only occur during night time and early hours of the morning. At higher RH ranges, the model continues to use the PBL height to make predictions, however, its reliance on the feature is less prevalent. This suggests that the model needs to rely on additional features to predict high RH ranges as opposed to mid-range RH where it can make a prediction primarily relying on the PBL height. For low RH levels (0-20\%), the model attributes very low weight on average to PBL height (illustrated as blue circles in Figure \ref{fig:vsn-results}\textbf{c}), which is possibly due to the fact that  samples with very low RH are present across all PBL heights. 

The next contributor to the model output predictions is the convective wind speed. This input feature contributes the most to predictions within the lower RH groups (0-40\% RH) as shown in Figure \ref{fig:vsn-results}\textbf{a}. The variation of model weights across wind speed values is depicted in Figure \ref{fig:vsn-results}\textbf{d}. From this figure, we can observe that the model heavily relies on convective wind speeds when the values of RH are low and wind speeds are higher (indicated in red). For such samples, the model relies on wind speeds for up to 83\% of the output prediction. This model behavior can be attributed to the low variation of RH values at higher wind speeds, hence the model has learned that when convective wind speeds are higher, the RH level is mostly likely lower. In contrast, at lower wind speeds, RH levels can vary significantly rendering these low wind speeds unreliable predictors of RH. The outcome is that the model assigns much lower weights to wind speeds when they are low (indicated in blue and green in Figure \ref{fig:vsn-results}\textbf{d}). 

Figure \ref{fig:vsn-results}\textbf{e} shows the model weights assigned to solar zenith angle (SZA) inputs as a function of the predicted RH and the SZA input values. The corresponding time of day is indicated below the angles to present a more intuitive understanding of the diurnal variation of model weights and RH predictions. The results show that the model relies more heavily on SZA when predicting high RH levels, particularly in the hours between midnight and sunrise, during which RH peaks. This model behavior could be attributed to the absence of higher RH levels beyond this range of solar zenith angles. Consequently, the model learns to rely on SZA as a form of ``condition'' to predict higher RH levels. The hours between sunrise and sunset (SZA = -90 to 90) are a strong indication of lower RH levels, although the model weights are extremely low for SZA within this interval (indicated by blue circles). This behavior could stem from the model effectively using alternative variables, such as convective wind speeds, to achieve robust predictions of lower RH levels, thereby reducing the need to rely on SZA. 

Other input features are minimally considered by the model to deduce RH, which is a useful outcome that would enable us to return to our input feature space and eliminate irrelevant variables, improving the model's computational efficiency both during training and inference. 

\section{Discussion and Conclusions}
\label{sec:discussion}

We propose a machine learning model that accurately predicts RH in Gale Crater by capturing complex relationships between RH and various meteorological variables from the Mars PCM. In particular, we train our model on both the MSL Curiosity rover's RH measurements as well as simulated meteorological variables that are generated by the Mars PCM as inputs in order to predict RH. This hybrid approach allows us to leverage both the increasing amount of observational data in addition to existing physical models of Mars, where it has proven to be effective in modeling RH as indicated by an $R^2$ coefficient of 0.92. While the model’s predictions were more accurate when predicting lower RH ranges (0-40\%) with its accuracy decreasing as RH rises, this trend aligns with rover measurement uncertainties, which rise from 2\% RH at 0\% RH to 20\% RH at 100\% RH. Importantly, the neural network architecture presented in this work is designed with interpretability in mind, facilitating a comprehensive understanding of its decision-making process which is a crucial feature when relying on its outcomes for further scientific research or mission planning. Our interpretation analysis reveals the key meteorological features from the Mars PCM that contribute significantly to our model's predictions; these features include the monthly mean surface H$_2$O layer, PBL height, convective wind speed, and solar zenith angle. We find that the model heavily relies on convective wind speeds to predict lower RH ($<$ 10\%) levels, minimally relying on other features. Conversely, to predict higher RH levels ($>$ 20\%), the model heavily utilizes all of the features mentioned above with its reliance on each feature varying across different samples. Furthermore, we train the VSN architecture using an alternative loss function, quantile loss, which outputs prediction intervals of RH. Here, we find that the prediction intervals were approximately similar to the uncertainty present in the rover RH measurements, suggesting that the model error is largely due to instrument uncertainty. 

The utilization of machine learning models such as the one presented in this study can contribute to a number of tasks. For instance, the model can fill temporal gaps in rover measurements by generating accurate synthetic data. This is important for analyses that require complete time series data or when comparing data obtained from orbiters and other rovers that may be taking measurements at different times. Comparison studies between different missions are essential for honing recalibration processes and gaining a more comprehensive view of how atmospheric data couples with the Martian surface. The operation of instruments on board rovers is often restricted to specific times of day, resulting in temporally limited data coverage. Machine learning modeling offers a practical solution to expand such data coverage and obtain measurements spanning any temporal window. Since the inputs to the machine learning model are generated by a physical model (Mars PCM), it is possible to generate these inputs for time periods that are lacking rover measurements. By feeding these inputs into the deep neural network, we can obtain accurate RH predictions, thereby expanding the temporal coverage of observations. Likewise, spatial gaps in observations around the Curiosity rover location can also be filled. 


RH prediction studies on Earth are largely tailored to the specific geographical locations in question. Similarly, the model proposed in this study can only be applied to model RH at Gale Crater surrounding the Curiosity rover location. The direct applicability of this particular trained model for RH prediction in diverse geographic locations on Mars is constrained due to variations in the interplay between RH near the surface and other meteorological variables across different regions. However, this neural network architecture can be applied to diverse geographical locations by training new instances of the model on different datasets that are representative of said locations. To enhance the capacity of this method, several prospective improvements could be pursued. Firstly, this method can be expanded to model other atmospheric variables such as temperature and wind in addition to RH and can be trained with data from any Mars mission, rovers or orbiters. Furthermore, the integration of time series forecasting techniques at multiple temporal scales could be explored. Currently, the model does not receive RH values as inputs. However, by providing past RH values as inputs, the model can be expanded to forecast future RH levels with higher precision. Lastly, the scope of this analysis can be expanded and performance may be improved by using two-dimensional spatial grids of physically modeled meteorological variables to predict two-dimensional spatial grids of RH in addition to other observed climate variables on Mars.

\section*{Acknowledgements}
This work was supported by the New York University Abu Dhabi
(NYUAD) Institute Research Grant G1502 and the ASPIRE Award
for Research Excellence (AARE) Grant S1560 by the Advanced
Technology Research Council (ATRC). This work utilized the High
Performance Computing (HPC) resources of NYUAD. We thank Prof. K. R. Sreenivasan for his constant encouragement and support for the project.

\section*{Author Contribution and Data Availability}
N.A., D.B.D., and D.A. designed the research. G.M. processed MSL data. N.A. analyzed the data. N.A. wrote the manuscript with contributions from D.B.D., D.A., and G.M. MSL data used for this manuscript is available at NASA PDS (Planetary Data System).


\bibliography{humidity}{}
\bibliographystyle{aasjournal}



\end{document}